\documentclass[prl,aps,twocolumn]{revtex4}
\usepackage{amsmath,graphicx,epsfig}

\usepackage{euscript}
\usepackage{amsfonts}
\usepackage{amssymb}
\usepackage{float}

\def\prn#1{{\left(#1\right)}}

\def\sbrk#1{{\left[#1\right]}}
\def\abrk#1{{\langle#1\rangle}}
\def\ket#1{{|#1\rangle}}
\def\bra#1{{\langle#1|}}

\def\cg(#1,#2)(#3,#4)(#5,#6){\bra{#1,#2,#3,#4}#5,#6\rangle}

\def\threej(#1,#2)(#3,#4)(#5,#6){\begin{pmatrix}#1&#3&#5\\#2&#4&#6\end{pmatrix}}
\def\sixj(#1,#2,#3)(#4,#5,#6){\begin{Bmatrix}#1&#2&#3\\#4&#5&#6\end{Bmatrix}}
\def\ninej(#1,#2,#3)(#4,#5,#6)(#7,#8,#9){\begin{Bmatrix}#1&#2&#3\\#4&#5&#6\\#7&#8&#9\end{Bmatrix}}

\def\sA{{\ensuremath{\EuScript A}}}

\def\sR{{\ensuremath{\EuScript R}}}

\def\sV{{\ensuremath{\EuScript V}}}

\def\mb{\mathbf}
\def\bs{\boldsymbol}

\newlength{\defbaselineskip}
\newcommand{\setlinespacing}[1]%
           {\setlength{\baselineskip}{#1 \defbaselineskip}}

\begin{document}

\title{Nuclear spin content and constraints on exotic spin-dependent couplings} 

\author{D. F. Jackson Kimball}
\email{derek.jacksonkimball@csueastbay.edu}
\affiliation{Department of Physics, California State University --
East Bay, Hayward, California 94542-3084, USA}

\date{\today}



\begin{abstract}
There are numerous recent and ongoing experiments employing a variety of atomic species to search for couplings of atomic spins to exotic fields. In order to meaningfully compare these experimental results, the coupling of the exotic field to the atomic spin must be interpreted in terms of the coupling to electron, proton, and neutron spins. Traditionally, constraints from atomic experiments on exotic couplings to neutron and proton spins have been derived using the single-particle Schmidt model for nuclear spin. In this model, particular atomic species are sensitive to either neutron or proton spin couplings, but not both. More recently, semi-empirical models employing nuclear magnetic moment data have been used to derive new constraints for non-valence nucleons. However, comparison of such semi-empirical models to detailed large-scale nuclear shell model calculations and analysis of known physical effects in nuclei show that existing semi-empirical models cannot reliably be used to predict the spin polarization of non-valence nucleons. The results of our re-analysis of nuclear spin content are applied to searches for exotic long-range monopole-dipole and dipole-dipole couplings of nuclei leading to significant revisions of some published constraints.
\end{abstract}



\maketitle

\section{Introduction}

There are numerous recent and ongoing atomic physics experiments searching for exotic spin-dependent couplings (see, for example, Refs.~\cite{Hun13,Hec08,Hec13,Vas09,Led13,Smi11,Kim13,Tul13,Bul13} and also Chapter 18 of Ref.~\cite{Bud13book} for a review). Such couplings are generated in a wide variety of theories postulating new physics beyond the Standard Model: for example, theories incorporating new scalar/pseudoscalar or axial vector interactions \cite{Moo84,Dob06,Fla09}, long-range torsion gravity \cite{Nev80,Nev82,Car94}, violation of Lorentz and CPT symmetries \cite{Kos11}, spontaneous breaking of Lorentz symmetry \cite{Ark05}, unparticles \cite{Geo07,Lia07}, and so on.

Experimental searches for exotic spin-dependent interactions utilize a wide variety of atoms and nuclei. In order to compare experimental sensitivities to new physics it is essential to determine how constraints on atomic spin-dependent interactions relate to constraints on atomic constituents. Here we address this essential question: how are experimental constraints on exotic spin-dependent couplings of atoms related to constraints on protons, neutrons, and electrons?

In particular, the most difficult problem is the relationship of constraints on exotic spin-dependent couplings of various nuclei to constraints on constituent nucleons, which requires input from nuclear theory. For the most part, experiments have been interpreted using the nuclear shell model \cite{May49,Jen49,Hax49} in the single-particle approximation (the Schmidt model \cite{Sch37}), but more recently semi-empirical models relying on measured nuclear magnetic moments have been widely applied \cite{Eng89,Fla06,Fla09,Ber11,Sta15}. One of the central points of the present work is that semi-empirical models have only limited applicability and can in many cases lead to inaccurate results, in particular for estimates of the proton(neutron) spin contribution to the nuclear spin of nuclei with even numbers of protons(neutrons). Where available, detailed nuclear shell model calculations can be applied to give more accurate estimates of nuclear spin content (see, for example, Refs.~\cite{Ots01,Cau05,Koo97} for reviews). In the course of our survey of nuclear spin content and the reliability of semi-empirical estimates, we derive new constraints on exotic long-range monopole-dipole and dipole-dipole interactions.

\section{Exotic atomic dipole moments}

We parameterize the spin couplings to new physics in terms of an exotic atomic dipole moment $\bs{\chi}=\chi_a\mb{F}$ related to coupling constants $\chi_e$, $\chi_p$, and $\chi_n$ for the electron, proton, and neutron, respectively (it is generally assumed that such couplings do not follow the same scaling as magnetic moments). In the following we also assume the new physics does not couple to orbital angular momentum. The nucleon coupling constants $\chi_p$ and $\chi_n$ can in turn be related to quark and gluon couplings via measurements and calculations based on quantum chromodynamics \cite{Aid13,Fla03,Fla04}.

The relationship of the expectation value for total atomic angular momentum $\abrk{ \mb{F} }$ to electron spin $\abrk{ \mb{S}_e }$ and nuclear spin $\abrk{ \mb{I} }$ can be reliably estimated for the ground states of most low-to-intermediate mass atoms based on the Russell-Saunders {\it{LS}}-coupling scheme:
\begin{align}
\abrk{ \mb{F} } &= \abrk{ \mb{S}_e } + \abrk{ \mb{L} } + \abrk{ \mb{I} }~, \nonumber \\
&= \frac{\abrk{ \mb{S}_e\cdot\mb{F} } }{ F(F+1) } \abrk{ \mb{F} } + \frac{\abrk{ \mb{L}\cdot\mb{F} }}{ F(F+1) } \abrk{ \mb{F} } + \frac{\abrk{ \mb{I}\cdot\mb{F} }}{ F(F+1) } \abrk{ \mb{F} }~,
\end{align}
where $\mb{L}$ is the orbital angular momentum. It follows that for the exotic atomic dipole moment coupling constant $\chi_a$,
\begin{align}
\chi_a = \chi_e\frac{\abrk{ \mb{S}_e\cdot\mb{F} }}{ F(F+1) } + \chi_N\frac{\abrk{ \mb{I}\cdot\mb{F} }}{ F(F+1) }~,
\label{Eq:exotic-atomic-dipole}
\end{align}
where $\chi_N$ is the exotic nuclear dipole coupling constant which can be expressed in terms of $\chi_p$ and $\chi_n$.

The projection of $\mb{S}_e$ on $\mb{F}$ can be calculated in terms of eigenvalues of the system according to:
\begin{widetext}
\begin{align}
\abrk{ \mb{S}_e\cdot\mb{F} } & = \frac{\abrk{ \mb{S}_e\cdot\mb{J} }}{J(J+1)} \abrk{ \mb{J}\cdot\mb{F} }~, \\
& = \frac{\sbrk{J(J+1)+S_e(S_e+1)-L(L+1)} \sbrk{F(F+1)+J(J+1)-I(I+1)}}{4J(J+1)}~,
\end{align}
\end{widetext}
where $\mb{J} = \mb{S}_e + \mb{L}$, and the projection of $\mb{I}$ on $\mb{F}$ is given by
\begin{align}
\abrk{ \mb{I}\cdot\mb{F} }  = \frac{1}{2}\sbrk{F(F+1) + I(I+1) - J(J+1)}~.
\end{align}

The next problem is a more difficult one: what is the relationship between $\chi_N$ and the nucleon coupling constants, $\chi_p$ and $\chi_n$? As noted in the introduction, a first estimate can be obtained from the nuclear shell model for odd-$A$ nuclei \cite{Kli52} by assuming that the nuclear spin $\mb{I}$ is due to the orbital motion and intrinsic spin of one nucleon only and that the spin and orbital angular momenta of all other nucleons sum to zero \cite{Bla79}. This is the assumption of the Schmidt or single-particle model \cite{Sch37}.  In the Schmidt model the nuclear spin $\mb{I}$ is generated by a combination of the valence nucleon spin ($\mb{S}_p$ or $\mb{S}_n$) and the valence nucleon orbital angular momentum $\bs{\ell}$, so that we have
\begin{align}
\chi_N & = \frac{\abrk{ \mb{S}_{p,n} \cdot \mb{I} }}{I(I+1)}\chi_{p,n}~, \\
& = \frac{S_{p,n}(S_{p,n}+1) + I(I+1) - \ell(\ell+1)}{2I(I+1)} \chi_{p,n}~, \label{Eq:Schmidt-model-chiN}
\end{align}
where it is assumed that the valence nucleon is in a well-defined state of $\ell$ and $S_{p,n}$.  To date, most atomic experiments searching for spin-dependent interactions have employed the Schmidt model for interpretation of their results.

However, it is well known that nuclear magnetic moments are only partially predicted by the Schmidt model, since in most cases it is a considerable oversimplification of the nucleus. Thus, in general, the nuclear spin content and magnetic moment cannot be described by a single valence nucleon in a well-defined state of $\ell$ and $S_{p,n}$.

One approach to understanding the spin content of nuclei is the application of rather complex and sophisticated large-scale nuclear shell model calculations, which have been carried out for a number of nuclei \cite{Cau05}.  These shell model calculations can be compared to a wide variety of experimental data: energy level spectra, electromagnetic moments (magnetic dipole, electric quadrupole, magnetic octopole, etc.), transition rates, and atomic hyperfine structure, which collectively inform the accuracy of the models.  Clearly, where available, these calculations are the most reliable model of nuclear spin content presently available.  However, there are many nuclei of interest for which detailed shell model calculations to predict spin content have not yet been carried out, and so there have been several attempts to predict nuclear spin content using semi-empirical estimates \cite{Eng89,Fla06,Fla09,Ber11,Sta15} in order to interpret experiments.

\section{Semi-empirical models based on magnetic moments}

In general, there are four different contributions to nuclear spin: the intrinsic spin of protons and neutrons, $\sigma_p\mb{I} = \abrk{\mb{S}_{p}}$ and $\sigma_n\mb{I} = \abrk{\mb{S}_{n}}$, respectively, as well as the orbital angular momentum of protons and neutrons, $\sigma_{\ell p}\mb{I} = \abrk{ \bs{\ell}_p }$ and $\sigma_{\ell n}\mb{I} = \abrk{ \bs{\ell}_n }$, respectively, where:
\begin{align}
\sigma_p + \sigma_n + \sigma_{\ell p} + \sigma_{\ell n} = 1~.
\label{Eq:spin-content-sum}
\end{align}
The exotic dipole moment coupling constant $\chi_N$ is related to $\chi_p$ and $\chi_n$ via:
\begin{align}
\chi_N = \chi_n \sigma_n  + \chi_p \sigma_p~.
\end{align}
The nuclear magnetic dipole moment, or, equivalently, the nuclear g-factor $g_I$, can be ascribed to the intrinsic magnetic moments of the nucleons as well as their orbital motion:
\begin{align}
g_I = g_p\sigma_p + g_n\sigma_n + g_{\ell p}\sigma_{\ell p} + g_{\ell n}\sigma_{\ell n}~,
\label{Eq:mag-moment-sum}
\end{align}
where $g_p$ and $g_n$ are the spin g-factors for the proton and neutron, respectively, and $g_{\ell p}$ and $g_{\ell n}$ are their orbital g-factors. For bare nucleons $g_p \approx 5.586$, $g_n \approx -3.826$, $g_{\ell p}=1$, and $g_{\ell n}=0$ \cite{Bla79}. It is apparent that if nuclear spin content is to be determined from the empirical values of the nuclear spin and magnetic moment, additional assumptions beyond Eqs.~\eqref{Eq:spin-content-sum} and \eqref{Eq:mag-moment-sum} are required.

In the Schmidt model for odd-$A$ nuclei, the nuclear ground state properties are determined by a single valence nucleon (either a proton or neutron: $p,n$), and thus
\begin{align}
1 & = \sigma_{s} + \sigma_{\ell}~, \label{Eq:single-particle-sum-relation1} \\
g_I & = g_{s}\sigma_{s} + g_{\ell}\sigma_{\ell}~,
\label{Eq:single-particle-sum-relation2}
\end{align}
where $s$ refers to the spin of the valence nucleon and $\ell$ refers to its orbital angular momentum.
Standard angular momentum relations for a state with total angular momentum = $I$, spin = $1/2$, and orbital angular momentum = $\ell$  [Eq.~\eqref{Eq:Schmidt-model-chiN}] give zeroth-order approximations for the $I=\ell + 1/2$ case:
\begin{align}
\sigma_{s}^{(0)} &= \frac{1}{2\ell + 1}~,\\
\sigma_{\ell}^{(0)} &= \frac{2\ell}{2\ell + 1}~,
\end{align}
and for the $I=\ell - 1/2$ case:
\begin{align}
\sigma_{s}^{(0)} &= -\frac{1}{2\ell + 1}~,\\
\sigma_{\ell}^{(0)} &= \frac{2\ell+2}{2\ell + 1}~,
\end{align}
which using Eq.~\eqref{Eq:single-particle-sum-relation2} yield the Schmidt model predictions for nuclear g-factors \cite{Bli56}.

Alternatively, Eqs.~\eqref{Eq:single-particle-sum-relation1} and \eqref{Eq:single-particle-sum-relation2} can be combined with the experimentally measured $g_I$ and the bare nucleon g-factors to extract a semi-empirical prediction for the spin content, which is the approach of Engel and Vogel \cite{Eng89} (hereafter denoted the EV model):
\begin{align}
\sigma_{s} &= \frac{g_I - g_{\ell}}{g_{s} - g_{\ell}}~,\\
\sigma_{\ell} &= \frac{g_I - g_s}{g_{\ell} - g_{s}}~.
\end{align}
The EV model is based on the assumption that the even system of nucleons (protons in odd-neutron nuclei and vice versa) carries little orbital or spin angular momentum.

Flambaum and Tedesco \cite{Fla06} introduced an alternative semi-empirical model (hereafter denoted the FT model) based on entirely different set of assumptions: the spin-orbit interaction is neglected and it is assumed that inter-nucleon forces conserve the total spin of neutrons and protons separately from the total orbital angular momentum, such that
\begin{align}
\sigma_{s}^{(0)} &= \sigma_p + \sigma_n~, \label{Eq:FT-spin-sum}\\
\sigma_{\ell}^{(0)} &= \sigma_{\ell p} + \sigma_{\ell n}~.
\end{align}
In the ``minimal'' FT model \cite{Fla06,Fla09,Ber11,Sta15}, changes to the orbital angular momenta predicted by the Schmidt model are neglected, so for nuclei with valence neutrons
\begin{align}
\sigma_p &= \frac{g_I - g_n \sigma_s^{(0)}}{g_p - g_n}~,
\end{align}
and for nuclei with valence protons
\begin{align}
\sigma_p &= \frac{g_I - \sigma_{\ell}^{(0)} - g_n \sigma_s^{(0)}}{g_p - g_n}~,
\end{align}
and $\sigma_n$ can be obtained from Eq.~\eqref{Eq:FT-spin-sum}. In the ``preferred'' FT model \cite{Fla06,Fla09,Ber11,Sta15}, it is assumed that the total angular momentum of protons is conserved separately from neutrons, so that for valence protons $\sigma_p~+~\sigma_{\ell p} = 1$, yielding
\begin{align}
\sigma_p = \frac{g_I - g_n \sigma_s^{(0)} - 1}{g_p - g_n - 1}~,
\end{align}
and for valence neutrons $\sigma_p + \sigma_{\ell p} = 0$,
\begin{align}
\sigma_p = \frac{g_I - g_n \sigma_s^{(0)}}{g_p - g_n - 1}~,
\end{align}
and again $\sigma_n$ can be obtained from Eq.~\eqref{Eq:FT-spin-sum}.

The critical question in regards to the use of these semi-empirical models to constrain exotic spin-dependent interactions of protons and neutrons is the accuracy of the calculated values of $\sigma_n$ and $\sigma_p$.  In the next section, we consider the application of these semi-empirical models to different nuclei of interest to assess their reliability.

\section{Case studies}

\subsection{$^{3}$He}

A well-studied case is $^3$He \cite{Gib92}, for which there have been both experimental determinations \cite{Ant96} and detailed shell model calculations (see, for example, Refs.~\cite{Fri88,Fri90,Bis01,Eth13} and references therein) of the contribution of neutron and proton spin polarization to the total spin of the $^3$He nucleus. Furthermore, $^3$H is a mirror nucleus of $^3$He, which allows comparisons that aid the understanding of the nuclear structure considerably \cite{Bli56,Gib92}. The ground state wave function is predominantly $^2S_{1/2}$.

The measurement of deep inelastic scattering of polarized electrons from a spin-polarized $^3$He target reported in Ref.~\cite{Ant96} found the neutron spin polarizations to be in excellent agreement with the calculations of Refs.~\cite{Fri88,Fri90,Bis01,Eth13}, and explains the departure from the expectation of the Schmidt model primarily as the result of two important factors: configuration mixing \cite{Ari54} that admixes a small amount of the $^4D_{1/2}$ state into the ground state and exchange effects related to virtual meson currents that quench the effective magnetic moments of the nucleons within the nucleus \cite{Vil47,Dre60,Ris72,Nob81}. Configuration mixing is known to be especially important in nuclei with nominal $s_{1/2}$ ground states because of the relatively strong mixing with $d$-states \cite{Ari54,Bli56}.

\begin{table}
\caption{Comparison of nuclear spin content and magnetic moment for $^3$He as predicted by the single-particle Schmidt model \cite{Sch37}, the semi-empirical EV model \cite{Eng89}, the minimal and preferred semi-empirical FT models \cite{Fla06}, and full-scale shell model calculations \cite{Fri90,Eth13}. For brevity only the total fraction of nuclear spin arising from orbital angular momentum, $\sigma_\ell = \sigma_{\ell n} + \sigma_{\ell p}$ is listed. The superscript $*$ indicates that the magnetic moment from experiment \cite{Flo93} has been used as input. For simplicity all calculations have been truncated after three decimal places, except the results from Ref.~\cite{Fri90} where uncertainties were determined by comparison to experimental data.}
\medskip \begin{tabular}{lcccc} \hline \hline
\rule{0ex}{3.6ex} Model~~~~ & ~~~$\sigma_n$~~~ &  ~~~$\sigma_p$~~~ & ~~~$\sigma_{\ell}$~~~ & ~~~$g_I$~~~ \\
\hline
\rule{0ex}{3.6ex} Schmidt  \cite{Sch37} &               1.000 & 0.000 & 0.000 & -3.826 \\
\rule{0ex}{3.6ex} EV  \cite{Eng89} &                    1.112 & 0.000 & -0.112 & -4.255$^*$ \\
\rule{0ex}{3.6ex} FT minimal  \cite{Fla06} &            1.046 & -0.046 & 0.000 & -4.255$^*$ \\
\rule{0ex}{3.6ex} FT preferred  \cite{Fla06} &          1.051 & -0.051 & 0.000 & -4.255$^*$ \\
\rule{0ex}{3.6ex} Mirror nuclei comp. \cite{Eng89} &    1.044 & -0.162 & 0.118 & 0.261$^*$ \\
\rule{0ex}{3.6ex} Full-scale shell  \cite{Fri90} &      0.87(2) & -0.027(4) & 0.16(2) & -4.220 \\
\rule{0ex}{3.6ex} Full-scale shell \cite{Eth13} &       0.856 & -0.029 & 0.173 & -- \\
\hline \hline
\end{tabular}
\label{Table:He-3}
\end{table}

Table~\ref{Table:He-3} compares the results from various calculations and highlights key problems with the semi-empirical models. Both versions of the FT model underestimate the importance of orbital angular momentum of nucleons, since spin-orbit coupling (which evidently plays an important role in nuclear structure \cite{Bla79}) is neglected. The EV model neglects polarization of the even system of nucleons (protons in the case of $^3$He), since it ascribes the difference between the Schmidt model prediction of $g_I$ and the experimental value entirely to configuration mixing within the odd system of nuclei.

Both empirical models systematically overestimate the contribution of neutron spin to the nuclear spin, which is related to the fact that neither of the semi-empirical models discussed in the previous section account for exchange effects. In fact, Engel and Vogel \cite{Eng89} were aware of the importance of exchange effects, and used experimental information from mirror nuclei to refine their model to include exchange effects for $^3$He and obtained the results shown in the fifth row of Table~\ref{Table:He-3}, which, however, still substantially disagree with the full-scale shell model calculations.

\subsection{$^{9}$Be}

In the case of $^9$Be, according to the shell model in the $jj$-coupling scheme \cite{Kli52}, there are 2 protons in the unfilled $p_{3/2}$ shell and 3 neutrons in the unfilled $p_{3/2}$ shell, while similarly in the $LS$-coupling scheme the nucleons occupy the $^2$P$_{3/2}$ ground state \cite{Ing53}.  The ground states of light nuclei tend to be better approximated by the $LS$-coupling scheme (in which single-particle spin-orbit coupling is neglected) rather than by the $jj$-coupling scheme (in which inter-nucleon forces are neglected), but $^9$Be is already within the intermediate coupling regime \cite{Bli56,Ing53,Fre55}. Consequently, configuration mixing is important.

Because of strong inter-nucleon forces, in cases like $^9$Be where there are multiple nucleons outside closed shells, the single-particle approximation of the Schmidt model which ascribes the ground-state nucleon properties to a single valence nucleon is often too extreme and a less restrictive assumption, referred to as the individual particle model \cite{Bli56,Miz51,Miz52}, is more appropriate. In the individual particle model, given an even number of protons(neutrons) $N_e$ and an odd number of neutrons(protons) $N_o$ in equivalent states outside closed shells with $N_o > N_e$, the ground state with isotopic spin $T=\prn{ N_e-N_o }/2$ has a probability of a proton occupying the valence state of
\begin{align}
\beta \approx \frac{N_e}{ \prn{2I+1}\prn{2T+2}  }
\end{align}
and a probability of the neutron occupying the valence state of $\alpha = 1-\beta$.
If $N_o < N_e$,
\begin{align}
\beta \approx \frac{2I+1 - N_e}{ \prn{2I+2}\prn{2T+2}  }~.
\end{align}
Note that the individual particle model prediction for $g_I$ of $^9$Be is in good agreement with the experimental value \cite{LBLnuclearmoments}.

\begin{table}
\caption{Comparison of nuclear spin content and magnetic moment for $^9$Be as predicted by the single-particle Schmidt model \cite{Sch37}, the individual particle model \cite{Miz52}, the semi-empirical EV model \cite{Eng89}, the minimal and preferred semi-empirical FT models \cite{Fla06}, and a full-scale shell model calculation \cite{Ara96}.}
\medskip \begin{tabular}{lcccc} \hline \hline
\rule{0ex}{3.6ex} Model~~~~ & ~~~$\sigma_n$~~~ &  ~~~$\sigma_p$~~~ & ~~~$\sigma_{\ell}$~~~ & ~~~$g_I$~~~ \\
\hline
\rule{0ex}{3.6ex} Schmidt  \cite{Sch37} &               0.333 & 0.000 & 0.667 & -1.275 \\
\rule{0ex}{3.6ex} Individual particle  \cite{Miz52} &   0.289 & 0.044 & 0.667 & -0.768 \\
\rule{0ex}{3.6ex} EV  \cite{Eng89} &                    0.205 & 0.000 & 0.795 & -0.785$^*$ \\
\rule{0ex}{3.6ex} FT minimal  \cite{Fla06} &            0.281 & 0.052 & 0.667 & -0.785$^*$ \\
\rule{0ex}{3.6ex} FT preferred  \cite{Fla06} &          0.275 & 0.058 & 0.667 & -0.785$^*$ \\
\rule{0ex}{3.6ex} Full-scale shell \cite{Ara96} &       0.238 & 0.000 & 0.762 & -0.779 \\
\hline \hline
\end{tabular}
\label{Table:Be-9}
\end{table}

Detailed shell-model calculations for $^9$Be have also been carried out \cite{Ara96,For05}, and point to a very different picture of the $^9$Be nucleus (which, in fact, was already basically understood in the 1950's \cite{Ing53}): the core of the nucleus is best described as a pair of $\alpha$ particles ($^4$He nuclei) and there is a valence neutron outside the core. The core possesses some orbital angular momentum and thereby affects $g_I$ through the orbital motion of the protons (contributing about +0.19 to $g_I$), but the only appreciable contribution from intrinsic spin comes from the valence neutron (whose state is affected by configuration mixing). As can be seen from Table~\ref{Table:Be-9}, in the case of $^9$Be, the EV model gives better agreement with the full-scale shell model calculation than either version of the FT model or the individual particle model, since core polarization of proton spin is found to be negligible in this particular case.

\subsection{$^{39}$K}

For nuclei that are a single nucleon away from a doubly closed shell, meaning that the shell is full in both the $jj$-coupling scheme and the $LS$-coupling scheme, configuration mixing should be minimal and the magnetic moment should be well-described by the Schmidt model. A good example is $^{17}$O which has 8 protons (a magic number) and 9 neutrons (one away from a magic number), with the valence neutron in a $d_{5/2}$ state, whose measured magnetic moment \cite{LBLnuclearmoments} is within 1\% of the Schmidt model value.

This is also the case for $^{39}$K, which has 20 neutrons (a magic number) and 19 protons (one away from a magic number). The valence proton (or, perhaps, proton ``hole'' in this case) is in a $d_{3/2}$ state. However, the agreement between the Schmidt model value and the measured magnetic moment is considerably worse in the case of $^{39}$K as compared to $^{17}$O. The main cause of this disagreement is that for odd-proton nuclei the magnetic moment is modified by velocity-dependent forces related to spin-orbit coupling \cite{Jen52,Bli55}: in the presence of an electromagnetic field the proton momentum $\mb{p}$ must be replaced by $\mb{p} - e \mb{A}(\mb{r})/c$, where $\mb{A}(\mb{r})$ is the vector potential at the position $\mb{r}$ of the proton, $e$ is the proton charge, and $c$ is the speed of light. In the case of $^{39}$K this effect shifts $g_I$ from the Schmidt value of $g_I=0.083$ to $g_I \approx 0.223$ \cite{Jen52}, in much better agreement with the experimental value of $g_I = 0.261$ \cite{LBLnuclearmoments}. This correction does not apply to valence neutrons, which explains why the Schmidt model's magnetic moment prediction agrees so well with experiment for $^{17}$O.  None of the semi-empirical models discussed account for this effect, which can have a dramatic impact on their estimates for odd-proton nuclei. Further corrections to the $^{39}$K magnetic moment due to relativistic effects and meson exchange effects \cite{Dre60} yield the estimate $g_I \approx 0.243$, within $\approx 7\%$ of the experimental value.

\begin{table}
\caption{Comparison of nuclear spin content and magnetic moment for $^{39}$K as predicted by the single-particle Schmidt model \cite{Sch37} with and without spin-orbit, relativistic, and exchange corrections to the magnetic moment \cite{Dre60}, the semi-empirical EV model \cite{Eng89}, the minimal and preferred semi-empirical FT models \cite{Fla06}, an estimate from comparison with mirror nuclei properties \cite{Eng89}, and a detailed perturbation theory calculation \cite{Tow87} carried out with two different spin-dependent residual interaction potentials, labeled I and II \cite{Eng05}.}
\medskip \begin{tabular}{lcccc} \hline \hline
\rule{0ex}{3.6ex} Model~~~~ & ~~~$\sigma_n$~~~ &  ~~~$\sigma_p$~~~ & ~~~$\sigma_{\ell}$~~~ & ~~~$g_I$~~~ \\
\hline
\rule{0ex}{3.6ex} Schmidt  \cite{Sch37} &                   0.000 & -0.200 & 1.200 & 0.083 \\
\rule{0ex}{3.6ex} Schmidt ($\mu$ corr.) \cite{Dre60} &      0.000 & -0.200 & 1.200 & 0.243 \\
\rule{0ex}{3.6ex} EV  \cite{Eng89} &                        0.000 & -0.161 & 1.161 & 0.261$^*$ \\
\rule{0ex}{3.6ex} FT minimal  \cite{Fla06} &               -0.019 & -0.181 & 1.200 & 0.261$^*$ \\
\rule{0ex}{3.6ex} FT preferred  \cite{Fla06} &             -0.021 & -0.179 & 1.200 & 0.261$^*$ \\
\rule{0ex}{3.6ex} Mirror nuclei comp. \cite{Eng89} &        0.037 & -0.131 & 1.094 & 0.261$^*$ \\
\rule{0ex}{3.6ex} Perturbation I \cite{Eng05} &             0.034 & -0.131 & 1.097 & 0.280 \\
\rule{0ex}{3.6ex} Perturbation II \cite{Eng05} &            0.036 & -0.123 & 1.087 & 0.124 \\
\hline \hline
\end{tabular}
\label{Table:K-39}
\end{table}

In addition to the EV and FT models, another semi-empirical estimate can be obtained from the mirror nuclei comparison carried out by Engel and Vogel \cite{Eng89}, which should account for exchange effects and correction to the proton magnetic moment reasonably well. Table~\ref{Table:K-39} compares Schmidt model results, semi-empirical estimates, and results of a detailed perturbation-theory calculation \cite{Eng05} based on the methods of Towner \cite{Tow87}.  The estimates for $\sigma_p$ are in relatively good agreement (within $\approx 50\%$ of one another) but the $\sigma_n$ estimates agree poorly with the perturbation theory calculations and the mirror nuclei comparison estimate.

\subsection{$^{129}$Xe and $^{131}$Xe}

\begin{table*}
\caption{Predictions of the fractional contribution of neutron and proton spins ($\sigma_n$ and $\sigma_p$, respectively) and neutron and proton orbital angular momentum ($\sigma_{\ell p}$ and  $\sigma_{\ell n}$, respectively) to the nuclear spin  of $^{129}$Xe and $^{131}$Xe for the Schmidt model \cite{Sch37}, semi-empirical models \cite{Eng89,Fla06}, and from detailed large-scale shell model calculations \cite{Res97,Toi09,Men12,Klo13,Vie15} (Bonn A and Nij. II refer to two different inter-nucleon force models employed in Ref.~\cite{Res97}). Also listed for comparison are calculated values of the nuclear g-factors (the superscript symbol $*$ indicates that the nuclear g-factors were input as constraints from experiment in the semi-empirical calculations).}
\medskip \begin{tabular}{lccccc|ccccc} \hline \hline
\rule{0ex}{3.6ex} ~ & \multicolumn{5}{c}{$^{129}$Xe} & \multicolumn{5}{c}{$^{131}$Xe} \\
\rule{0ex}{2.6ex} ~~~~~~~~~~ & ~~~~$\sigma_n$~~~~ & ~~~~$\sigma_p$~~~~ & ~~~~$\sigma_{\ell n}$~~~~ & ~~~~$\sigma_{\ell p}$~~~~ & ~~~~$g_I$~~~~ & ~~~~$\sigma_n$~~~~ & ~~~~$\sigma_p$~~~~ & ~~~~$\sigma_{\ell n}$~~~~ & ~~~~$\sigma_{\ell p}$~~~~ & ~~~~$g_I$~~~~ \\
\hline
\rule{0ex}{3.6ex} Schmidt  \cite{Sch37}                         & 1.000 & 0.000 & 0.000 & 0.000 & -3.826 & -0.200 & 0.000 & 1.200 & 0.000 & 0.765 \\
\rule{0ex}{3.6ex} EV  \cite{Eng89}                              & 0.407 & 0.000 & 0.593 & 0.000 & -1.556$^*$ & -0.120 & 0.000 & 1.120 & 0.000 & 0.461$^*$ \\
\rule{0ex}{3.6ex} FT minimal  \cite{Fla06}                      & 0.759 & 0.241 & 0.000 & 0.000 & -1.556$^*$ & -0.168 & -0.032 & 1.200 & 0.000 & 0.461$^*$ \\
\rule{0ex}{3.6ex} FT preferred  \cite{Fla06}                    & 0.730 & 0.270 & 0.000 & 0.000 & -1.556$^*$ & -0.164 & -0.036 & 1.200 & 0.000 & 0.461$^*$ \\
\rule{0ex}{3.6ex} Shell (Bonn A)  \cite{Res97}                  & 0.718 & 0.056 & -0.228 & 0.454 & -1.966 & -0.151 & -0.006 & 1.048 &  0.110 & 0.653 \\
\rule{0ex}{3.6ex} Shell (Nij. II) \cite{Res97}                  & 0.600 & 0.026 & -0.370 & 0.744 & -1.402 & -0.145 & -0.008 & 1.009 &  0.143 & 0.653 \\
\rule{0ex}{3.6ex} Shell \cite{Toi09}                            & 0.546 & -0.004 & 0.226 & 0.230 & -1.600 & -0.083 & -0.0005 & 0.945 & 0.139 & 0.453 \\
\rule{0ex}{3.6ex} Shell \cite{Men12,Klo13,Vie15}                & 0.658 & 0.020 & -0.644 & 0.966 & -1.440 & -0.181 & -0.006 & 1.273 & -0.086 & 0.573 \\
\hline \hline
\end{tabular}
\label{Table:Xe-comparisons}
\end{table*}

There have been a number of detailed, large-scale shell model calculations of $\sigma_n$, $\sigma_p$, $\sigma_{\ell p}$, and  $\sigma_{\ell n}$ for $^{129}$Xe and $^{131}$Xe \cite{Res97,Toi09,Men12,Klo13,Vie15} (see Table~\ref{Table:Xe-comparisons}) which can be compared to the semi-empirical models. This comparison for these relatively heavy nuclei with considerable configuration mixing again highlights the shortcomings of the semi-empirical models: the essential point is that the assumptions used in the semi-empirical models are not generally satisfied. The assumption of the EV model that restricts configuration mixing to system of odd nucleons is not satisfied: the xenon isotopes exhibit significant mixing between neutron and proton states. The FT model assumption of negligible spin-orbit coupling is also clearly not satisfied in the xenon isotopes: in $^{129}$Xe for example, configuration mixing generates a very significant contribution of neutron and proton orbital angular momentum to the nuclear spin, whereas the FT models assume the contribution is zero.  In particular we would like to draw attention to the fact that none of the semi-empirical models predict any contribution of orbital angular momentum of the even system of nucleons ($\sigma_{\ell p}$ in the case of $^{129}$Xe and $^{131}$Xe) to the nuclear spin, whereas all the shell models predict a relatively large $\sigma_{\ell p}$. As a result, the semi-empirical predictions for $\sigma_p$ in $^{129}$Xe and $^{131}$Xe are grossly inaccurate, and it can be inferred that they cannot reliably be used for prediction of $\sigma_p$ in heavy odd-neutron nuclei (nor for prediction of $\sigma_n$ in heavy odd-proton nuclei).

\subsection{$^{133}$Cs}

The valence proton of $^{133}$Cs is nominally in a $g_{7/2}$ state, but considerable configuration mixing is expected \cite{Ari54,Bli56}, along with renormalization of magnetic moments due to velocity-dependent forces \cite{Jen52,Bli55} (as discussed in regards to the case of $^{39}$K). Large-scale shell model calculations have been performed for $^{133}$Cs \cite{Toi09,Iac91} and are compared to the semi-empirical model predictions in Table~\ref{Table:Cs-133}.  Predictions of the Schmidt, semi-empirical, and shell model calculations are in reasonable agreement for the values of $\sigma_p$ and $\sigma_\ell$, but again there is significant disagreement in predictions of the spin polarization of the even system of nucleons, $\sigma_n$.

\begin{table}
\caption{Comparison of nuclear spin content and magnetic moment for $^{133}$Cs as predicted by the single-particle Schmidt model \cite{Sch37}, the semi-empirical EV model \cite{Eng89}, the minimal and preferred semi-empirical FT models \cite{Fla06}, and a detailed large-scale shell-model calculation \cite{Toi09}.}
\medskip \begin{tabular}{lcccc} \hline \hline
\rule{0ex}{3.6ex} Model~~~~ & ~~~$\sigma_n$~~~ &  ~~~$\sigma_p$~~~ & ~~~$\sigma_{\ell}$~~~ & ~~~$g_I$~~~ \\
\hline
\rule{0ex}{3.6ex} Schmidt  \cite{Sch37} &                   0.000 & -0.111 & 1.111 & 0.490 \\
\rule{0ex}{3.6ex} EV  \cite{Eng89} &                        0.000 & -0.057 & 1.057 & 0.738$^*$ \\
\rule{0ex}{3.6ex} FT minimal  \cite{Fla06} &               -0.026 & -0.085 & 1.200 & 0.738$^*$ \\
\rule{0ex}{3.6ex} FT preferred  \cite{Fla06} &             -0.029 & -0.082 & 1.200 & 0.738$^*$ \\
\rule{0ex}{3.6ex} Large-scale shell \cite{Toi09} &          0.006 & -0.091 & 1.135 & 0.820 \\
\rule{0ex}{3.6ex} Large-scale shell \cite{Iac91} &          0.0006 & -0.064 & 1.063 & -- \\
\hline \hline
\end{tabular}
\label{Table:Cs-133}
\end{table}

\subsection{$^{199}$Hg and $^{201}$Hg}

\begin{table*}
\caption{Comparison of nuclear spin content and magnetic moments for $^{199}$Hg and $^{201}$Hg as predicted by the single-particle Schmidt model \cite{Sch37}, the semi-empirical EV model \cite{Eng89}, and the minimal and preferred semi-empirical FT models \cite{Fla06}.}
\medskip \begin{tabular}{lccccc|ccccc} \hline \hline
\rule{0ex}{3.6ex} ~ & \multicolumn{5}{c}{$^{199}$Hg} & \multicolumn{5}{c}{$^{201}$Hg} \\
\rule{0ex}{2.6ex} ~~~~~~~~~~ & ~~~~$\sigma_n$~~~~ & ~~~~$\sigma_p$~~~~ & ~~~~$\sigma_{\ell n}$~~~~ & ~~~~$\sigma_{\ell p}$~~~~ & ~~~~$g_I$~~~~ & ~~~~$\sigma_n$~~~~ & ~~~~$\sigma_p$~~~~ & ~~~~$\sigma_{\ell n}$~~~~ & ~~~~$\sigma_{\ell p}$~~~~ & ~~~~$g_I$~~~~ \\
\hline
\rule{0ex}{3.6ex} Schmidt  \cite{Sch37}                         & -0.333 & 0.000 & 1.333 & 0.000 & 1.275 & 0.333 & 0.000 & 0.667 & 0.000 & -1.275 \\
\rule{0ex}{3.6ex} EV  \cite{Eng89}                              & -0.265 & 0.000 & 1.265 & 0.000 & 1.012$^*$ & 0.097 & 0.000 & 0.903 & 0.000 & -0.373$^*$ \\
\rule{0ex}{3.6ex} FT minimal  \cite{Fla06}                      & -0.305 & -0.028 & 1.333 & 0.000 & 1.012$^*$ & 0.237 & 0.096 & 0.667 & 0.000 & -0.373$^*$ \\
\rule{0ex}{3.6ex} FT preferred  \cite{Fla06}                    & -0.302 & -0.031 & 1.333 & 0.000 & 1.012$^*$ & 0.226 & 0.107 & 0.667 & 0.000 & -0.373$^*$ \\
\hline \hline
\end{tabular}
\label{Table:Hg-comparisons}
\end{table*}

To our knowledge, there are no large-scale shell model calculations for $^{199}$Hg or $^{201}$Hg, atoms of considerable interest for use in searches for exotic spin-dependent interactions \cite{Hun13,Ven92,You96}. The valence neutron of $^{199}$Hg is in a $p_{1/2}$ state, for which configuration mixing coupling constants are nominally zero \cite{Ari54,Bli56}, so the Schmidt model is expected to give a reasonable approximation of the nuclear spin content. Indeed, the Schmidt model prediction for the magnetic moment is much closer to the experimental value for $^{199}$Hg (within $\approx 20\%$) than for $^{129}$Xe, $^{131}$Xe, or $^{201}$Hg. Meson-exchange effects are expected to contribute only a small correction to the $^{199}$Hg magnetic moment \cite{Dre60,Ris72,Nob81}, so most likely the residual discrepancy between the Schmidt model value for $g_I$ and the experimental value is due to higher-order configuration mixing terms. Although the semi-empirical model predictions for $\sigma_n$ are all relatively close to one another (Table~\ref{Table:Hg-comparisons}), as we have seen in the examples discussed above, the semi-empirical models cannot accurately ascertain the relative contribution of proton spin ($\sigma_p$) and orbital motion ($\sigma_{\ell p}$ and $\sigma_{\ell n}$). The valence neutron of $^{201}$Hg is nominally in a $p_{3/2}$ state and so significant configuration mixing is expected \cite{Ari54,Bli56}, and once again the semi-empirical models are in stark disagreement as to corrections to the Schmidt predictions for spin content.

\section{Exotic long-range monopole-dipole couplings}

To extract constraints on exotic spin-dependent interactions from recent searches it is also crucial to examine the details of the measured experimental signature. As a first example, we re-analyze constraints on long-range monopole-dipole couplings. Assuming the monopole-dipole coupling originates from one-boson exchange within a Lorentz-invariant quantum field theory, a light scalar/pseudoscalar field generates a monopole-dipole potential $\sV_{9,10}(r)$ of the form (the subscript is in reference to enumerated potentials in Ref.~\cite{Dob06}):
\begin{align}
\sV_{9,10}(r) = \frac{{\rm g}_p^X{\rm g}_s^Y\hbar}{8\pi m_Xc} \mb{S}_X \cdot \hat{\mb{r}} \prn{ \frac{1}{r\lambda} + \frac{1}{r^2} }  e^{-r/\lambda} ~,
\label{Eq:monopole-dipole-potential}
\end{align}
where ${\rm g}_p^X$ is the dimensionless pseudoscalar coupling constant for particle $X$, ${\rm g}_s^Y$ is the dimensionless scalar coupling constant for particle $Y$, $m_X$ is the mass of particle $X$, $\mb{r}$ is the displacement vector between $X$ and $Y$, $\lambda$ is the range of the new force, $\hbar$ is Planck's constant, $c$ is the speed of light, and $\mb{S}_X$ is the intrinsic spin of particle $X$ in units of $\hbar$. Assuming a long-range force (communicated by a massless boson) we may approximate $\lambda \rightarrow \infty$, which gives for the monopole-dipole potential:
\begin{align}
\sV_{9,10}(r) = \frac{{\rm g}_p^X{\rm g}_s^Y\hbar}{8\pi m_Xc r^2} \mb{S}_X \cdot \hat{\mb{r}}~.
\label{Eq:monopole-dipole-potential-infinite-range}
\end{align}

If the new scalar/pseudoscalar field is considered to be an additional component of gravity, as suggested by certain scalar-tensor extensions of general relativity based on a Riemann-Cartan spacetime \cite{Heh76,Sha02,Ham02,Kos08}, the interaction could be considered a coupling of spins to gravitational fields.  The dominant gravitational field in a laboratory setting is that due to the Earth, which generates a spin-dependent Hamitonian with the nonrelativistic form \cite{Fla09}:
\begin{align}
H_g = k_X \frac{\hbar}{c} \bs{S}_X\cdot\bs{g}~,
\label{Eq:GDM-Hamiltonian}
\end{align}
where $k_X$ is a dimensionless parameter setting the scale of the new interaction for particle $X$ and $\bs{g}$ is the acceleration due to gravity.  If the strength of the pseudoscalar coupling is the same as that of the usual tensor component of gravity, $k_X \approx 1$ \cite{Fla09}, setting the scale for the energy difference between opposite spin orientations with respect to $\bs{g}$ at $\approx 4 \times 10^{-23}~{\rm eV}$ (corresponding to a spin precession frequency of $\approx 10^{-8}~{\rm Hz}$). The connection between Eqs.~\eqref{Eq:monopole-dipole-potential-infinite-range} and \eqref{Eq:GDM-Hamiltonian} is obtained by integrating the contribution of all the constituent particles making up the Earth ($\sim M_E/m_p$, where $M_E$ is the mass of the Earth and $m_p$ is the proton mass):
\begin{align}
k_X \approx \frac{{\rm g}_p^X{\rm g}_s}{8\pi m_X g R_E^2} \frac{M_E}{m_p}~,
\end{align}
where $R_E$ is the radius of the Earth. Note that in the above estimate there is an implicit assumption about scalar coupling to constituent particles in the Earth: we have assumed equal scalar coupling (${\rm g}_s$) to protons and neutrons, which we also assume to have nearly equal abundance in the Earth, and neglect scalar coupling to electrons. Of course other assumptions could be made that would change the extracted limits.

\begin{figure*}
\center
\includegraphics[width = 3.5 in]{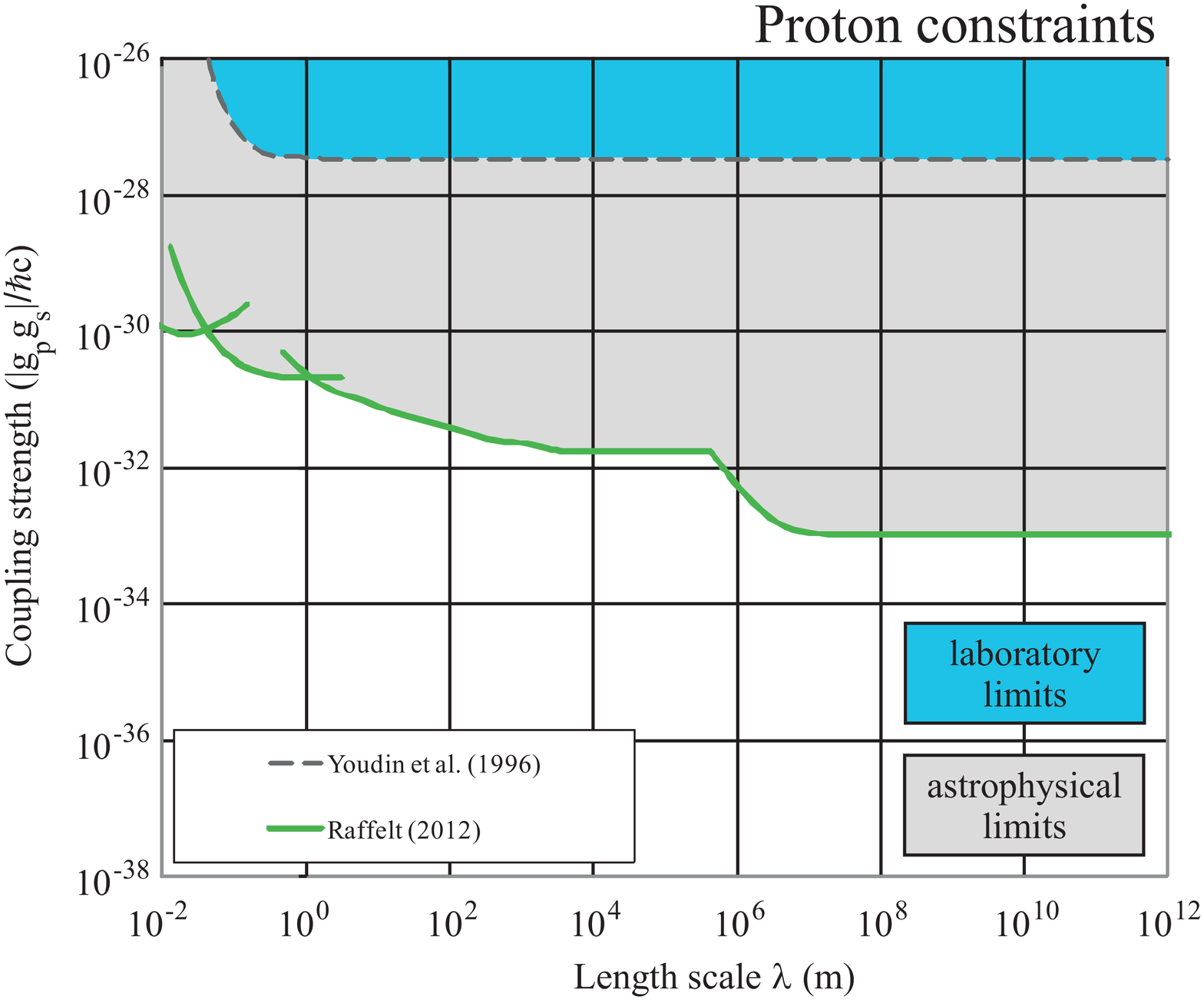}
\includegraphics[width = 3.5 in]{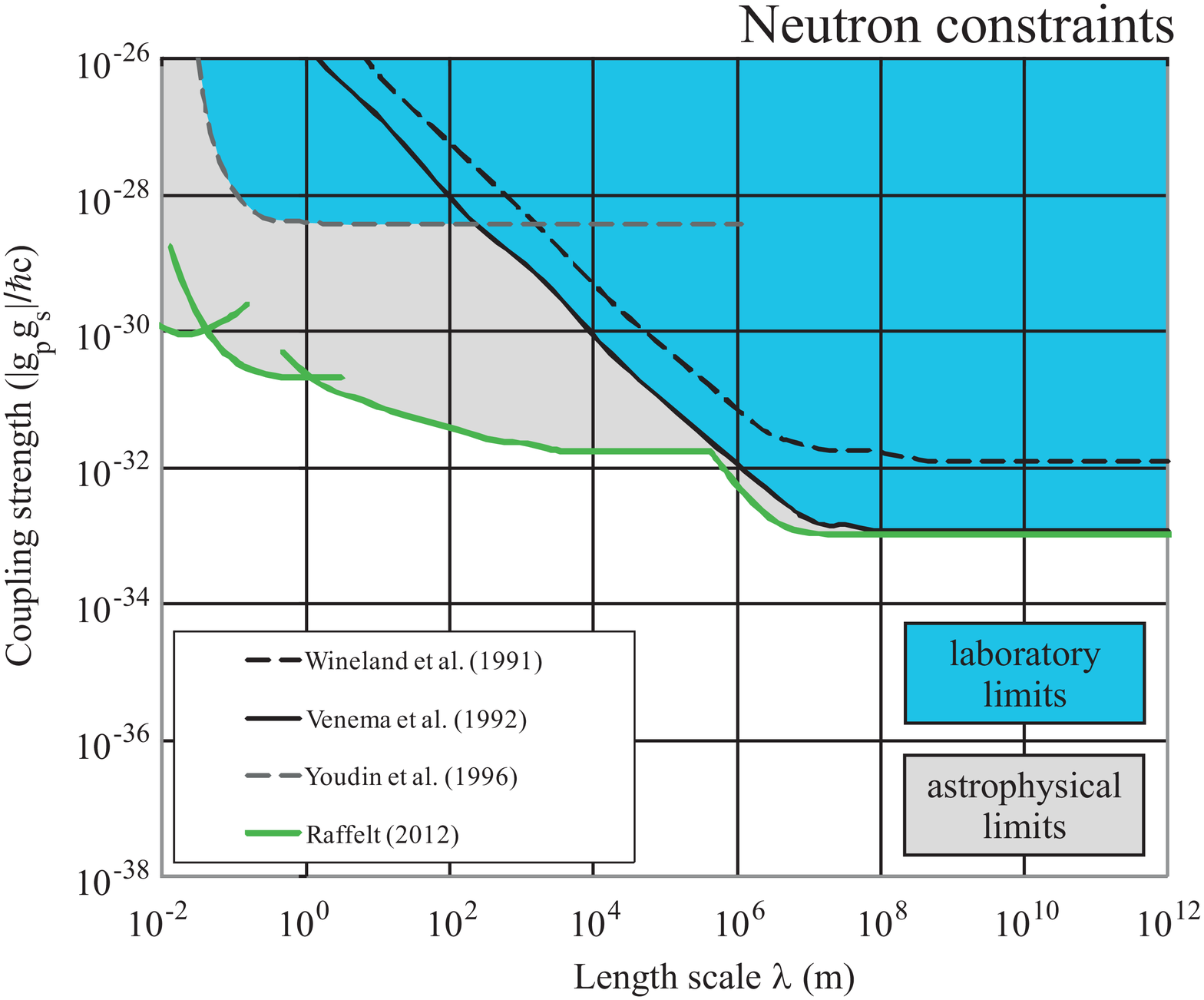}
\caption{Laboratory constraints (shaded blue, from Refs.~\cite{Win91,Ven92,You96}) on monopole-dipole (scalar-pseudoscalar) couplings between nucleons, $\left| {\rm g}_p {\rm g}_s \right|/\hbar c$, as a function of the range $\lambda$ of the interaction (${\rm g}_p$ and ${\rm g}_s$ are the pseudoscalar and scalar coupling constants, respectively). Astrophysical constraints for baryon couplings (excluded parameter space shaded grey) are from the recent analysis of Raffelt (2012) \cite{Raf12}. The plot on the left shows constraints on proton couplings, where presently only astrophysical bounds apply, and the plot on the right shows constraints on neutron couplings. Note that based on the re-analysis carried out in this work, this plot is significantly different from the plot published in Ref.~\cite{Kim13}.}
\label{Fig:monopole-dipole-parameter-exclusion-plot}
\end{figure*}

Presently the best constraints on long-range monopole-dipole couplings of nuclear spins are obtained from the experiment of Venema {\it et al.} \cite{Ven92} comparing spin precession of mercury isotopes ($^{199}$Hg and $^{201}$Hg) and the experiment of Wineland {\it et al.} \cite{Win91} measuring hyperfine transitions in $^9$Be$^+$ ions. Both experiments searched for the coupling of spins to the mass of the Earth. In the shell model $^9$Be, $^{199}$Hg and $^{201}$Hg all have valence neutrons, and so the above experiments are principally sensitive to neutron couplings (see Fig.~\ref{Fig:monopole-dipole-parameter-exclusion-plot}).

The experiment of Venema {\it et al.} \cite{Ven92} explicitly constrains the quantity
\begin{align}
\sA \epsilon' < 3.0 \times 10^{-21}~{\rm eV}~,
\end{align}
where $\sA$ is the strength of the monopole-dipole interaction,
\begin{align}
\sA \approx \frac{{\rm g}_p^X{\rm g}_s\hbar}{16\pi m_X c R_E^2} \frac{M_E}{m_p}  \approx  k_X \frac{\hbar g}{2c}~,
\end{align}
and
\begin{align}
\epsilon' = \epsilon_{201} - \frac{g_{201}}{g_{199}}\epsilon_{199}~,
\end{align}
where $g_{201}/g_{199} = -0.369139$ expresses the ratio of the Hg Land\'e g-factors and $\epsilon_{199}$ and $\epsilon_{201}$ parameterize the exotic spin coupling of the Hg nuclei.

The estimated values of $\sigma_n$ from the Schmidt model and semi-empirical estimates agree to within $\approx 20\%$ for $^{199}$Hg and to within a factor of 2 for $^{201}$Hg (Table~\ref{Table:Hg-comparisons}). Furthermore, from our survey of nuclear spin content in the previous section, we have found that the Schmidt model estimates of the contribution of the valence nucleon to the nuclear spin are generally within a factor of two of the results of large-scale shell structure calculations. Thus we estimate that $\sigma_n(^{199}{\rm Hg}) = -0.3(1)$ and $\sigma_n(^{201}{\rm Hg}) = +0.3(1)$, and from these values compute $\epsilon'$ for the neutron:
\begin{align}
\epsilon'_n = \sigma_n(^{201}{\rm Hg}) - \frac{g_{201}}{g_{199}} \sigma_n(^{199}{\rm Hg}) = 0.2(1)~,
\end{align}
The resulting constraint on long-range monopole-dipole couplings of the neutron is shown in Fig.~\ref{Fig:monopole-dipole-parameter-exclusion-plot}, and the limit on the spin-gravity coupling parameter $k_n$ is given by
\begin{align}
k_n \lesssim 10^3~.
\end{align}

The experiment of Wineland {\it et al.} \cite{Win91} constrains the frequency shift between the $^9$Be$^+$ $^2 S_{1/2}~\ket{F=1,M=0}$ and $^2 S_{1/2}~\ket{F=1,M=-1}$ states caused by a long-range monopole-dipole interaction. The expectation values of the nuclear and electron spin projections along the quantization axis for the $\ket{F=1,M=0}$ state are zero, and for the $\ket{F=1,M=-1}$ state:
\begin{align}
\abrk{ \sigma_e }_z &= \frac{1}{2}~, \\
\abrk{ \sigma_N }_z &= -\frac{5}{6}~.
\end{align}
The frequency shift when the leading magnetic field was reversed relative to $\bs{g}$ was constrained to be $< 13.4~\mu{\rm Hz}$ (or in energy units $< 5.5 \times 10^{-20}~{\rm eV}$). The full-scale shell-model calculation of Arai {\it et al.} \cite{Ara96} found that $\sigma_n \approx 0.24$ (Table~\ref{Table:Be-9}), which should be accurate to better than $\approx 10\%$, resulting in the constraint on long-range monopole-dipole couplings of the neutron shown in Fig.~\ref{Fig:monopole-dipole-parameter-exclusion-plot}, about an order of magnitude less stringent than that obtained from Ref.~\cite{Ven92}.

In the case of $^{199}$Hg and $^{201}$Hg, since no reliable estimate of $\sigma_p$ is available, no constraint on proton couplings can extracted from the measurements reported in Ref.~\cite{Ven92}. In the case of $^9$Be, the shell model calculation of Arai {\it et al.} \cite{Ara96}, along with the considerations of nuclear structure discussed in the previous section, indicate that $^9$Be likely has vanishingly small proton spin polarization, and so again no limits on proton couplings can be obtained.  This is in contrast to previous analysis \cite{Fla09,Kim13} that reported proton limits based on the FT model.

Constraints for the proton can be obtained from the experiment by Youdin {\it{et al.}} \cite{You96} that searched for laboratory-range monopole-dipole couplings between a 475-kg lead mass and the spins of $^{133}$Cs and $^{199}$Hg atoms. In Ref.~\cite{You96}, the results of the experiment were interpreted to constrain electron and neutron spin couplings. Since in the shell model $^{133}$Cs has a valence proton, the Schmidt model, semi-empirical models, and large-scale shell model calculations are all in reasonable agreement concerning the value of $\sigma_p$ for $^{133}$Cs.  Thus the experiment by Youdin {\it{et al.}} \cite{You96} establishes the first laboratory constraints on long-range monopole-dipole (spin-gravity) couplings of the proton:
\begin{align}
\frac{{\rm g}_p^p{\rm g}_s}{\hbar c} \lesssim 3 \times 10^{-28}
\end{align}
for a range of $\gtrsim 20$~cm, and
\begin{align}
k_p \lesssim 3 \times 10^8~.
\end{align}
The corresponding constraints are shown in Fig.~\ref{Fig:monopole-dipole-parameter-exclusion-plot}.

Constraints on long-range monopole-dipole couplings of neutrons and protons can be compared to recent experiments searching for long-range monopole-dipole couplings of electrons using a spin-polarized torsion pendulum \cite{Hec08}, which obtained the constraint $k_e < 10$ from searching for a $\mb{S}_e \cdot \bs{g}$ correlation. By searching for a long-range monopole-dipole interaction where the source mass was the sun, the constraint ${\rm g}_p^e{\rm g}_s/(\hbar c) < 2 \times 10^{-36}$ was established \cite{Hec08}.

The above analysis is of particular relevance to our ongoing experiment to search for a long-range monopole-dipole (spin-gravity) coupling of Rb spins to the mass of the Earth \cite{Kim13}. In our experiment, we measure the ratio of the difference between the $^{87}$Rb and $^{85}$Rb precession frequencies in the ground state $F=2$ and $F=3$ states divided by their sum,
\begin{align}
\sR = \frac{\Omega_{87}-\Omega_{85}}{\Omega_{87}+\Omega_{85}}~.
\end{align}
Measurement of the ratio $\sR$ eliminates or reduces several common-mode sources of noise and systematic error.  Taking the difference between $\sR$ for a leading magnetic field parallel with $\bs{g}$ and anti-parallel with $\bs{g}$ yields a signal proportional to the spin precession frequency caused by nonmagnetic interactions. Ultimately the sensitivity of the experiment to long-range monopole-dipole couplings, $\delta k$, is related to the sensitivity to anomalous frequency shifts, $\delta \Omega$. The sensitivity of the experiment to proton couplings can be estimated based on the $\sigma_p$ value from the Schmidt model, taking into account the fractional contribution of the nuclear spin to the total atomic spin given by Eq.~\eqref{Eq:exotic-atomic-dipole}. The valence nucleons of $^{87}$Rb and $^{85}$Rb are protons, so the Schmidt model's estimate of $\sigma_p$ should be reliable to within a factor of two, yielding
\begin{align}
\delta k_p &\approx 3 \times 10^2 \times \delta \Omega ({\mu \rm Hz}) ~,
\end{align}
If our experiment can achieve sub-$\mu$Hz sensitivity to anomalous frequency shifts, our sensitivity to proton spin-gravity couplings would exceed existing laboratory constraints by six orders of magnitude.  Note that because of these reconsiderations of the nuclear spin content, the updated parameter exclusion plot shown in Fig.~\ref{Fig:monopole-dipole-parameter-exclusion-plot} is significantly different from the parameter exclusion plot published as Fig.~1 in Ref.~\cite{Kim13}.

While in this section we have focused our attention on the case of long-range monopole-dipole interactions, we note that there are also a number of experiments that search for short-range monopole-dipole interactions using a variety of experimental methods: for example, the study of spin-relaxation of polarized $^3$He \cite{Pet10}, nuclear magnetic resonance studies with $^{129}$Xe and $^{131}$Xe \cite{Bul13} and $^3$He \cite{Chu13}, and measurements of ultracold neutrons \cite{Ser10,Jen14}. None of the constraints listed are affected by our analysis, since they either used neutrons directly or reliable shell-model calculations for their estimates of nuclear spin content.

\section{Exotic long-range dipole-dipole couplings}

As a second example, we consider a long-range dipole-dipole coupling $\sV_3(r)$ between nuclei, which can be generated by a pseudoscalar field (again the subscript is in reference to enumerated potentials in Ref.~\cite{Dob06}):
\begin{widetext}
\begin{equation}
\sV_3(r) = \frac{{\rm g}_p^X{\rm g}_p^Y\hbar^2}{16\pi m_Xm_Y c^2}  \sbrk{ \mb{S}_X \cdot \mb{S}_Y \prn{ \frac{1}{\lambda r^2} + \frac{1}{r^3} }
 - \prn{ \mb{S}_X\cdot\hat{\mb{r}} } \prn{ \mb{S}_Y\cdot\hat{\mb{r}} } \prn{ \frac{1}{\lambda^2r} + \frac{3}{\lambda r^2} + \frac{3}{r^3}} } e^{-r/\lambda}~.
\label{Eq:dipole-dipole-potential}
\end{equation}
\end{widetext}
If the exchange boson is assumed to be nearly massless, the range of the interaction $\lambda \rightarrow \infty$ and we obtain
\begin{equation}
\sV_3(r) = \frac{{\rm g}_p^X{\rm g}_p^Y\hbar^2}{16\pi m_Xm_Y c^2r^3}  \sbrk{ \mb{S}_X \cdot \mb{S}_Y
 - 3\prn{ \mb{S}_X\cdot\hat{\mb{r}} } \prn{ \mb{S}_Y\cdot\hat{\mb{r}} } }~.
\label{Eq:dipole-dipole-potential-infinite-range}
\end{equation}

\begin{table*}
\caption{Constraints on long-range dipole-dipole couplings of neutrons and protons, using the parameterization of Eq.~\eqref{Eq:dipole-dipole-potential-infinite-range}. Constraints on proton spin couplings based on $^{129}$Xe are re-interpreted using the latest large-scale shell model calculations of $\sigma_n$ and $\sigma_p$ \cite{Men12,Klo13,Vie15}. For calculations related to $^3$He, the values of $\sigma_p$ and $\sigma_n$ from Ref.~\cite{Fri90}, reliable to within a few percent, are used.}
\medskip \begin{tabular}{lccc} \hline \hline
\rule{0ex}{3.6ex} System [Ref.]~~~~~ & ~~~~~~${\rm g}_p^n{\rm g}_p^n/(4\pi\hbar c)$~~~~~~ & ~~~~~~${\rm g}_p^n{\rm g}_p^p/(4\pi\hbar c)$~~~~~~ & ~~~~~~${\rm g}_p^p{\rm g}_p^p/(4\pi\hbar c)$~~~~~~\\
\hline
\rule{0ex}{3.6ex} $^{3}$He~/~$^{129}$Xe \cite{Gle08} & $< 3.5 \times 10^{-7}$  & $< 1.4 \times 10^{-6}$ & $< 3.1 \times 10^{-3}$ \\
\rule{0ex}{3.6ex} $^{3}$He~/~$^{3}$He \cite{Vas09} & $< 5.8 \times 10^{-10}$  & $< 1.7 \times 10^{-8}$ & $< 4.9 \times 10^{-7}$ \\
\hline \hline
\end{tabular}
\label{Table:dipole-dipole-limits}
\end{table*}

There have been two recent experiments \cite{Vas09,Gle08} nominally searching for laboratory-scale dipole-dipole interactions between polarized neutrons.

The experiment of Glenday {\it et al.} \cite{Gle08} measured the spin precession frequencies of $^3$He and $^{129}$Xe in a dual-species maser as the polarization of a nearby dense $^3$He gas was reversed. The authors of Ref.~\cite{Gle08} assessed the nuclear spin content using the FT approach for $^{129}$Xe (which had been applied to $^{129}$Xe in an earlier work \cite{Dzu85}) and measurements of deep inelastic scattering of electrons for $^3$He \cite{Ant96}. The constraints can be re-assessed using the values of $\sigma_n$ and $\sigma_p$ from the latest, largest-scale shell-model calculations for $^{129}$Xe \cite{Men12,Klo13,Vie15}, leading to the limits listed in Table~\ref{Table:dipole-dipole-limits}.

In the experiment of Vasilakis {\it et al.} \cite{Vas09} a spin-exchange relaxation free (SERF) comagnetometer \cite{Kor02,Kor05} using $^{39}$K and $^{3}$He was used to search for exotic dipole-dipole couplings to a nearby dense $^3$He gas. In the analysis carried out in Ref.~\cite{Vas09}, only the $^3$He spins were considered and the analysis focused on neutron-neutron couplings (again using the nuclear spin content determined by Ref.~\cite{Ant96}). In the SERF regime, the valence electron of $^{39}$K is rapidly kicked between the ground state $F=2$ and $F=1$ hyperfine levels by spin-exchange collisions and therefore the effective fraction of atomic spin due to nuclear spin polarization is given by the weighted average of the nuclear spin polarization in the two ground state hyperfine levels: $\sigma_N = 5/8$. The SERF comagnetometer has similar sensitivity to magnetic field couplings for $^{39}$K and $^{3}$He (in the experiment described in Ref.~\cite{Vas09} the magnetometric sensitivity reached $\sim 0.5~{\rm aT}$). However, it is crucial to note that the similar magnetometric sensitivity translates into different energy sensitivities for the two species due to the different magnetic moments of $^{39}$K and $^{3}$He: the sensitivity of $^{39}$K to exotic spin couplings is reduced by a factor of $\sim 130$ as compared to $^{3}$He (the ratio of the magnetic moments taking into account the statistical weighting between $^{39}$K ground-state hyperfine levels due to the rapid spin-exchange collisions). Because of this factor, in spite of the larger $\sigma_p$ for $^{39}$K as compared to $^3$He, constraints from this experiment on exotic couplings to $^{3}$He spins are more sensitive to proton couplings than those derived from exotic couplings to $^{39}$K. The derived constraints on long-range dipole-dipole couplings for neutrons and protons are shown in Table~\ref{Table:dipole-dipole-limits}.
(It should be noted that in the re-interpretation of the constraints from the experiments of Glenday {\it et al.} \cite{Gle08} and Vasilakis {\it et al.} \cite{Vas09}, the original species analyzed in the experiments are used and as such a simple re-scaling from theory can be applied without re-examination of the statistical or systematic error analysis.)

To put these constraints into context, the previous best limit reported in the literature for a long-range exotic dipole-dipole coupling of the form expressed in Eq.~\eqref{Eq:dipole-dipole-potential-infinite-range} between protons was that obtained by Ramsey using spectroscopy of molecular hydrogen \cite{Ram79}: ${\rm g}_p^p{\rm g}_p^p/(4\pi\hbar c) < 2.3 \times 10^{-5}$. The results obtained by Vasilakis {\it et al.} \cite{Vas09} combined with re-analysis of the nuclear spin content improve these constraints by a factor of $\sim 50$. These new constraints on exotic proton dipole-dipole couplings are quite reliable since they are derived from the well-measured and understood $\sigma_p$ of $^3$He.  Note that the constraints on long-range exotic dipole couplings of electron spins are far more stringent: the spin-polarized torsion pendulum experiment of Ref.~\cite{Hec13} reports ${\rm g}_p^e{\rm g}_p^e/(4\pi\hbar c) < 2.2 \times 10^{-16}$.

\section{The Global Network of Optical Magnetometers to search for Exotic physics (GNOME)}

As a final point, we consider a new experimental effort being initiated to search for exotic spin-dependent interactions that produce transient signals \cite{Pos13,Pus13}. While a single comagnetometer system, such as the SERF comagnetometer described in Refs.~\cite{Kor02,Kor05,Vas09}, could detect such transient events, it would be exceedingly difficult to confidently distinguish a true signal generated by heretofore undiscovered physics from ``false positives'' induced by occasional abrupt changes of comagnetometer operational conditions (e.g., magnetic-field spikes, laser-mode jumps, electronic noise, etc.).  Effective vetoing of false positive events requires an array of comagnetometers.  Furthermore, there are key benefits in terms of noise suppression and event characterization to widely distributing the comagnetometers geographically.  The Laser Interferometer Gravitational Wave Observatory (LIGO) collaboration has developed sophisticated data analysis techniques \cite{LIGO04} to search for similar correlated ``burst'' signals from a worldwide network of gravitational wave detectors, and we have recently demonstrated that these data analysis techniques can be applied to data from synchronized comagnetometers \cite{Pus13}.  Our proposed comagnetometer array, the Global Network of Optical Magnetometers to search for Exotic physics (GNOME), would be uniquely sensitive, for example, to cosmic events generating coherent bursts or waves of a heretofore undiscovered field \cite{Car94}, to correlated noise produced by a fluctuating \cite{Ell04} or oscillating \cite{Bud14} background field whose time-averaged value is zero, or passage through topological defects such as pseudoscalar domain walls \cite{Pos13}.  Eventually, the GNOME will consist of at least five dedicated atomic comagnetometers located at geographically separated stations.

Construction and testing of a prototype sensor for the GNOME is presently underway. The design is based on the SERF comagnetometer scheme developed by Romalis, Kornack, and colleagues \cite{Kor02,Kor05,Vas09,Smi11}. Our prototype sensor will utilize coupled $^3$He and $^{87}$Rb spins, which, according to the analysis presented here, offers sensitivity to exotic electron, neutron, and proton spin couplings. Based on Eq.~\eqref{Eq:exotic-atomic-dipole} and the surrounding discussion combined with the values of $\sigma_p$ and $\sigma_n$ from the Schmidt model for $^{87}$Rb and the values of $\sigma_p$ and $\sigma_n$ for $^3$He \cite{Fri90}, the exotic dipole moments for $^3$He and $^{87}$Rb in terms of $\chi_e$, $\chi_n$, and $\chi_p$ are given by:
\begin{align}
\chi\prn{{\rm ^{87}Rb}; F=2} &= \frac{1}{4} \chi_e + \frac{3}{4}\chi_N \\
& = 0.25 \chi_e + 0.25 \chi_p ~,\\
\chi\prn{{\rm ^{87}Rb}; F=1} &= -\frac{1}{4} \chi_e + \frac{5}{4}\chi_N \\
& = -0.25 \chi_e + 0.42 \chi_p ~,\\
\chi\prn{{\rm ^3He}} &= 0.87 \chi_n - 0.03 \chi_p  ~.
\end{align}
Due to the averaging over $^{87}$Rb hyperfine levels due to rapid spin-exchange collisions in the SERF regime, the effective exotic atomic dipole moment for $^{87}$Rb is given by
\begin{align}
\chi\prn{{\rm ^{87}Rb}} = 0.06 \chi_e + 0.31 \chi_p ~.
\end{align}
As noted previously in our discussion of the SERF comagnetometer scheme, in order to determine the relative sensitivity of $^{87}$Rb and $^3$He to new physics the reduction in sensitivity of $^{87}$Rb relative to $^3$He by the ratio of the magnetic moments must also be taken into account. As a result, the GNOME will be most sensitive to exotic spin-dependent couplings to neutrons, with sensitivity to proton couplings (primarily arising from the $\sigma_p$ of $^3$He) reduced by a relative factor of $\sim 30$ and sensitivity to electron couplings reduced by a relative factor of $\approx 2000$.

\section{Conclusions}

In conclusion, we have analyzed nuclear spin content for several nuclei of interest in searches for exotic spin-dependent interactions and assessed the reliability of semi-empirical models \cite{Eng89,Fla06,Fla09,Ber11,Sta15} that employ nuclear magnetic moment data to derive spin content of nuclei. We find that, based on physical processes known to be important in nuclear structure and comparison to large-scale shell model calculations, the assumptions of the semi-empirical models are not generally satisfied and thus their predictions are inaccurate, especially for spins of non-valence nucleons. An essential problem for the semi-empirical models is systematic error in the estimation of nucleon orbital angular momentum.  Applying revised nuclear spin content to results of previous experiments, we have re-derived constraints on long-range monopole-dipole (spin-gravity) and dipole-dipole interactions. In particular, we have corrected previously reported constraints on long-range monopole-dipole couplings of the proton \cite{Fla09,Kim13} and established new laboratory bounds on exotic monopole-dipole couplings between protons at a range of $\gtrsim 20~{\rm cm}$ based on the work of Youdin {\it{et al.}} \cite{You96}. We have also found that the experiment of Vasilakis {\it{et al.}} \cite{Vas09} constrains long-range exotic dipole-dipole couplings between protons over an order of magnitude more stringently than previously reported in the literature. In general, the interpretation of experiments searching for exotic spin-dependent interactions of nuclei would greatly benefit from more detailed nuclear theory calculations that could more reliably predict nuclear spin content.

\acknowledgments

The author is sincerely grateful to Dmitry Budker, Maxim Pospelov, Volker Koch, Feng Yuan, Victor Flambaum, Yevgeniy Stadnik, and Michael Romalis for enlightening discussions.  This work was supported by the National Science Foundation under grant PHY-1307507.  Any opinions, findings and conclusions or recommendations expressed in this material are those of the authors and do not necessarily reflect those of the National Science Foundation. This research was also supported in part by the Perimeter Institute for Theoretical Physics. Research at Perimeter Institute is supported by the Government of Canada through Industry Canada and by the Province of Ontario through the Ministry of Economic Development and Innovation.

\end{document}